\documentclass[aps,prx,floats,superscriptaddress]{revtex4-2}

\usepackage{graphicx,epsfig}

\usepackage{fullpage}
\usepackage{booktabs}
\usepackage{graphics,dcolumn,bm,epic, eepic,fleqn,float,fontenc}
\usepackage{amssymb,amsmath,amsfonts,multirow,rotate,color}
\usepackage{lipsum}
\usepackage{soul,xcolor,url}

\usepackage[title]{appendix}
\usepackage{hyperref}
\hypersetup{
    colorlinks=true, 
    linkcolor=blue, 
    citecolor=blue, 
    urlcolor=blue 
}

\makeatletter
\renewcommand{\thetable}{\arabic{table}}
\makeatother

\begin{document}

\title{Cycling into the workshop: predictive maintenance for Barcelona's bike-sharing system}

\author{Jordi Grau-Escolano}
\affiliation{Eurecat, Centre Tecnològic de Catalunya}
\affiliation{Universitat Politècnica de Catalunya-BarcelonaTech}

\author{Aleix Bassolas}
\affiliation{Eurecat, Centre Tecnològic de Catalunya}

\author{Julian Vicens}
\affiliation{Eurecat, Centre Tecnològic de Catalunya}

\begin{abstract}
  Bike-sharing systems have emerged as a significant element of urban mobility, providing an environmentally friendly transportation alternative. With the increasing integration of electric bikes alongside mechanical bikes, it is crucial to illuminate distinct usage patterns and their impact on maintenance. Accordingly, this research aims to develop a comprehensive understanding of mobility dynamics, distinguishing between different mobility modes, and introducing a novel predictive maintenance system tailored for bikes. By utilising a combination of trip information and maintenance data from Barcelona's bike-sharing system, Bicing, this study conducts an extensive analysis of mobility patterns and their relationship to failures of bike components. To accurately predict maintenance needs for essential bike parts, this research delves into various mobility metrics and applies statistical and machine learning survival models, including deep learning models. Due to their complexity, and with the objective of bolstering confidence in the system's predictions, interpretability techniques explain the main predictors of maintenance needs. The analysis reveals marked differences in the usage patterns of mechanical bikes and electric bikes, with a growing user preference for the latter despite their extra costs. These differences in mobility were found to have a considerable impact on the maintenance needs within the bike-sharing system. Moreover, the predictive maintenance models proved effective in forecasting these maintenance needs, capable of operating across an entire bike fleet. Despite challenges such as approximated bike usage metrics and data imbalances, the study successfully showcases the feasibility of an accurate predictive maintenance system capable of improving operational costs, bike availability, and security.
\end{abstract}

\keywords{bike-sharing, predictive maintenance, survival analysis, forecasting}

\maketitle

\section{Introduction}

Bike-sharing systems (BSSs) are a crucial part of urban mobility solutions, providing an eco-friendly alternative to private vehicles. Users notably reduce their reliance on other transportation modes \cite{pbs_in_modal_shift}, leading to increased physical activity and travel time savings \cite{pbs_advantages}. Additionally, urban environments benefit from reduced fuel consumption and improved economic development \cite{pbs_advantages}. Moreover, BSSs have experienced substantial growth since the 2000s, with 2022's global landscape encompassing nearly 2,000 BSSs and an impressive fleet of approximately 9 million bicycles \cite{medin_PBS_report}. The scale of these systems and technological advancements have enabled the generation of large amounts of data, providing researchers and decision-makers with valuable insights to improve existing systems.

BSSs have been extensively studied from various angles, including station location optimization \cite{station_optimization} and fairness \cite{duran2020fair}, bike rebalancing strategies \cite{Chardon2016}, as well as studies on user behavior changes and their impact on health \cite{pbs_in_modal_shift, pbs_advantages}. While substantial research has focused on human mobility within BSSs in recent years \cite{Talavera_Garcia2021, Kon2021}, the advent of fourth-generation BSSs has introduced a new layer of complexity through the global adoption of electric bikes (e-bikes) \cite{Shaheen2010, medin_PBS_report}. Existing BSSs, such as Barcelona's Bicing, in Spain, have progressively integrated e-bikes into their mechanical bike (m-bike) fleets, while there are instances of entirely new fully electric systems like Barcelona metropolitan area's AMBici. As BSS users increasingly embrace e-bikes for their speed, convenience, and reduced physical exertion, significant changes in existing BSS mobility dynamics are expected. However, the evolving mobility dynamics resulting from the coexistence of both mechanical and electric transportation modes remain an area ripe for study.

Another significant aspect to consider in the context of BSSs is the maintenance operations (MOs) for bikes. While the importance of maintenance as a critical activity for all companies has been underscored in various studies \cite{CHO19911, Reinertsen1996ResidualLO}, BSS maintenance practices have predominantly relied on the traditional approach of applying corrective measures (i.e., addressing issues as they arise by replacing worn-out or damaged parts). Therefore, enhancing the BSS user experience and operational efficiency could be achieved by developing a predictive maintenance strategy capable of anticipating and preventing bike component failures, thereby improving bike availability and security while reducing costs.

To address these topics, this article leverages both trip and maintenance datasets provided by Bicing BSS. Our research revolves around four key questions. Firstly, we delve into whether m-bikes and e-bikes can be categorized as two distinct mobility modes, despite sharing the same infrastructure. This involves an in-depth analysis of their respective usage and mobility patterns, aiming to identify any significant disparities between the two. Secondly, we explore bike component failure patterns, seeking to understand how the bike model factor contributes to the wear and tear of distinct parts. Thirdly, we aim to assess the potential of forecasting failures in three specific bike components by evaluating the accuracy of various survival models. Lastly, we aim to identify the critical factors that influence predicting the longevity of the bike components.

\section{Background}

  \subsection{BSS mobility}
Research in BSS mobility typically involves the examination of aspects like temporal usage patterns and trip characteristics, which include factors such as distance, duration, and speed. This kind of analysis serve in multiple purposes, such as examine general system dynamics \cite{Jensen2010Characterizing, Ciancia2015Exploring, Zaltz2013Structure}, designing effective rebalancing strategies \cite{Chiariotti2018Rebalancing}, predicting BSS demand \cite{Borgnat2011SharedBI}, and estimating trip destinations and durations \cite{Zhang2016Bicycle}. While trip characteristics can vary depending on the specific city in which the BSS is implemented, existing publications generally concur on certain approximate values. On average, BSS trips tend to have a mean distance of between 1 and 2 kilometers, a mean duration ranging from 10 to 20 minutes, and an average speed in the range of 10 to 15 kilometers per hour. Regarding trip distances, even though GPS signal can be used to establish ride routes \cite{Talavera_Garcia2021}, when not available, shortest paths between the origin and destination stations are computed \cite{Oliveira2016}.

An alternative perspective for studying BSS mobility emphasizes the geographical aspects of mobility. While some studies have examined the overall usage of BSSs across multiple cities \cite{Zhao2014Ridership, Duran2019Built, Sarkar2015Comparing}, much of this research is centered around understanding the patterns within individual BSSs. Most previous studies in this domain focus on the usage of individual docking stations, often utilizing concepts such as incoming and outgoing trips \cite{Moncayo2016Visualization, Faghih2014Land, Wang2016Modeling, Faghih2017Empirical, Duran2019Built, Oliveira2016}, as well as station occupancy \cite{OBrien2014Mining, Sarkar2015Comparing}. However, for a more comprehensive understanding of BSS mobility structures, the flows between stations have also been explored \cite{Corcoran2014Spatiotemporal, Yang2019Spatiotemporal, Kon2021}.

A relatively less studied area is the relationship between urban topology and BSS usage, specifically how the altitude difference between the origin and destination stations of a trip influences bike usage. In this regard, \cite{Morency2015ModellingBU} found that elevation has a negative impact on the number of incoming trips and a positive impact on the outgoing, while \cite{Kim2020Anatomy} found that altitude difference, together with stations' distance and weather features are good predictors of bike usage.

On the other side, when characterizing BSS mobility, the majority of research has predominantly focused on traditional m-bikes, with only a limited number of studies comparing their usage with other transportation modes. For instance, \cite{Noussan2019Urban} analyzed the temporal variations in BSS and car usage, while \cite{Jensen2010Characterizing} compared BSS trip distances and average speeds with those of car and pedestrians. Moreover, there is an important field of investigations in comparing between e-bikes and scooters. Spatio-temporal mobility patterns of BSSs and dockless scooter-sharing services were compared in \cite{Mckenzie2019Spatiotemporal}, finding that BSSs are mainly used for commuting, whereas scooter-sharing serves different purposes. Additionally, \cite{Almannaa2020Comparative} explored the average speed of shared e-scooters and BSS e-bikes, finding that e-bikes generally travel faster than e-scooters.

Mobility data from Bicing has served as a valuable resource in various articles, with the majority of studies relying on data from around 2010, a period characterized by fewer stations, bikes, and bike lanes compared to the situation in 2022. Notably, this was also a time before the introduction of e-bikes into the Bicing system. In this context, two early studies obtained from the Bicing website the number of occupied and empty bike docks at stations with the purpose of unraveling the general spatio-temporal patterns of Barcelona dynamics \cite{froehlich2008measuring} and comparing BSS usage patterns on weekdays and weekends employing hierarchical clustering \cite{froehlich2009sensing}. Additionally, Bayesian networks were used to predict the number of available bicycles at each station. In the same line, \cite{kaltenbrunner2010urban} utilized the same data to generate bike availability predictions in the stations using Auto-Regressive Moving Average (ARMA) models. More recently, in 2022, a study delved into predicting the usage of the BSS system and examined the impact of the COVID-19 pandemic on these predictions \cite{Bustamante2022Riding}. This last research used a much more comprehensive and updated dataset, including data from 2020 and 2021, which contained information about the origin and destination stations, as well as the start and end times of individual trips.

  \subsection{Predictive maintenance}

Since the 1990s, the scientific literature has emphasized the importance of maintenance as a critical activity for companies to improve reliability and reduce costs \cite{CHO19911, Reinertsen1996ResidualLO}. Over time, knowledge and techniques have evolved, transitioning from a corrective maintenance, which involved addressing failures after they occurred, to a preventive maintenance \cite{ran2019survey}, which implements scheduled maintenance based on time intervals to prevent breakdowns. While this approach helps to mitigate most failures, it comes with the drawback of high prevention costs. With the increased computational power, the advancement of artificial intelligence, and the rise of the IoT, a new strategy called predictive maintenance (PM) has emerged, which can predict failures before they happen, allowing an even lower failure rate and reduced costs. One of its main challenges is fault prognosis, which focuses on forecasting when a failure is likely to occur \cite{wang2017paradigm}. By accurately predicting failures, maintenance activities can be scheduled in advance, minimizing downtime, and optimizing resource allocation.

Reliability theory plays a significant role in PM modelling, and it is essentially the same as survival analysis (SA) \cite{Yang2022}. SA was originally developed in the field of biomedical sciences to examine life tables \cite{cox1972regression}, but its concept of events can be applied to various domains, such as machine component failures. A key challenge in SA studies is censoring, which refers to missing data when an event is not observed \cite{leung1997censoring, gijbels2010censored}. Censoring occurs not due to technical failures but rather due to the nature of the studied event. For instance, if a participant in a clinical trial decides to discontinue his participation before the event of interest occurs. It is precisely the presence of censored data that makes impractical the application of predictive algorithms using the usual statistical and machine learning approaches.

According to \cite{wang2019machine}, survival methods can be classified into statistical and machine learning methods. Statistical methods focus on characterizing the distribution of event times and the statistical properties of parameter estimation, such as estimating survival curves. These first models can be further divided into: (1) Non-parametric models (Kaplan-Meier, Nelson-Aalen, Life-Table), which make no assumptions about the underlying distribution and estimate the survival curves directly from the data. (2) Semi-parametric models (Cox model, CoxBoost, Time-dependent Cox), which incorporate both non-parametric estimation of the baseline survival function and parametric estimation of the effects of covariates. (3) Parametric models (Penalized regression, Accelerated Failure Time), which assume a specific distribution for the survival time and estimate the parameters of that distribution. Machine learning methods combine traditional SA techniques with machine learning algorithms, such as survival trees, Bayesian methods, neural networks, or support vector machines. Advanced machine learning techniques, including ensemble learning, active learning, transfer learning, and multitask learning, have also been applied in the field of SA.

These methodologies have been applied in a wide range of fields, including healthcare \cite{Reddy2015ARO}, reliability \cite{Modarres2016Reliability}, crowdfunding \cite{Li2016Project}, student retention \cite{Ameri_2016Survival}, customer lifetime \cite{Furrer2022International}, and unemployment duration analysis \cite{Kiefer1988Economic}. However, to the best of our knowledge, there are no applications of SA to BSSs. Apart from SA, scientific literature on bike's PM is scarce. \cite{MountainBike2019} proposed using smartphone vibration readings and support vector machine models to predict the health of mountain bike’s components (the rotor, the chain, the wheel bearings, the steering head, and the derailleur cog). However, the scope of this study was not a fleet of bikes but a single bicycle. On the contrary, \cite{Predictivemaintenancebike}’s main objective was to study the cyclists’ behavioral patterns in the BSS of Oslo, Finland, and successfully identify the need for bike maintenance. In this case, random forest models were applied to the ride (destination, duration, and date) and the cyclist (gender and year of birth) data. Finally, \cite{Matkovic2021Towards} focuses on predicting brakes’ performance with KNN, LSTM and XGBoost classifiers, using as input physical influences and acceleration/deceleration forces coming from hall and inertial IoT sensors.

\section{Data and methods}

  \subsection{Case of study}

Launched in 2007, Bicing BSS operates in the city of Barcelona, Spain, with strategically located stations across the city that serve as docking points for bicycles. Over the years, Bicing has experienced multiple expansions, and, as of December 2022, it included nearly 150,000 unique users, approximately 7,000 bikes, and 519 permanent stations with between 12 and 54 docking points. One distinctive feature of this BSS is the presence of two types of bikes since 2019: e-bikes and m-bikes, with the presence of batteries that allow motorized assistance up to 25 km/h on e-bikes being the main difference between them. The number of e- and m-bikes have evolved over time. In 2019, e-bikes represented only 15\% of a fleet that comprised 6,700 bikes. Since then, there has been a gradual shift, with 2,000 m-bikes being upgraded to e-bikes, and additional e-bikes being introduced. As a result, in December 2022, e-bikes account for 47\% of the expanded fleet. The usage of these bikes is associated with an annual payment and, additionally, with fees that are typically based on the duration of the rental. Moreover, e-bikes are associated with a small initial cost for each ride.

  \subsection{Data}

To answer the previously exposed research questions, Bicing has made available to this article two bike-sharing data sets: a trips dataset and a maintenance dataset (Appendix \ref{app_data}).

\begin{itemize}
    \item The trips data set encompasses individual trips generated from April 2019 to December 2022 (i.e., 3 years and 9 months). Out of 53 million trips, 33 million (62\%) were completed by m-bikes, while 20 million (38\%) were made with e-bikes. Each trip entry includes the following information: starting and ending dates and times with second-level granularity, starting and ending stations, bike identifier, bike model (m-bike or e-bike), and an anonymized user identifier.
    \item Maintenance data is comprised of a total of 310,000 MOs, which correspond to various bicycle repairs types executed between September 1st, 2020, and January 1st, 2023 (i.e., 2 years and 4 months). Each MO record provides the following information: MO identifier, date, category, subcategory, bike identifier, and bike model (m-bike or e-bike).
    There exist a total of 12 categories and 87 subcategories, encompassing a wide array of actions such as cleaning, greasing, adjusting, or changing bike parts. Categories range from fewer than 100 interventions to 120,000, with the majority of them falling into the brake and wheel categories (66\%). Likewise, subcategories also reveal a wide range of counts, with only 10\% of the repair typologies surpassing 10,000 MOs.
  \end{itemize}
  
Additionally, geographical data for the 519 stations, including latitude and longitude coordinates, were provided for the study.
    
  \subsection{Trips processing} \label{section_trips_processing}
  
To study mobility patterns and develop the predictive maintenance strategy, only trips with durations ranging from 2 to 60 minutes were taken into account, which constituted 99\% of all trips. Once filtered, various trip metrics were collected for the mobility analysis. While trip duration could be directly derived from the trips data, other measures required some additional steps. Obtaining the trip routes was not feasible since data only provided the starting and ending stations for each trip. As a consequence, like in \cite{Oliveira2016}, we presumed that trips followed the shortest paths between stations, and these routes were obtained using the OpenStreetMaps API \cite{OpenStreetMap}. Then, using the trips' duration and distance their average speed was obtained. Furthermore, the cumulative trip inclinations were simplified by calculating the difference between the altitudes of the origin and destination stations. The specific station altitudes were obtained using the OpenTopoData API \cite{OpenTopoData}.

To determine whether the trip characteristics of m-bikes and e-bikes come from the same distribution, 5,000 samples from each subgroup were randomly selected. Initially, a Shapiro-Wilk test was applied to both samples to assess whether they followed a normal distribution. Subsequently, if both samples were found to be normally distributed, an Independent Sample T-Test was applied for comparison. In cases where one or both samples did not meet the normality assumption, the Kolmogorov-Smirnov test was used. All statistical tests were conducted with a significance level set at 0.01.

To compare trip numbers of both bike models, several metrics were
computed to avoid comparisons with absolute numbers. Initially, for each bike model, at all stations, the percentage of incoming (or outgoing) trips was determined by dividing the total trips for each model by the station’s total trips. Then,
two supplementary metrics were derived from these percentages: (1) the differences in the percentage of incoming (or outgoing) trips between e-bikes and m-bikes, and (2) the difference in the percentage of incoming and outgoing trips for each bike model.

  \subsection{MOs processing} \label{section_mos_processing}

First, the target bike parts for this study needed to be selected. To ensure complete objectivity, subjective MO types such as cleaning or greasing were excluded, and the focus was placed on the replacement of bike parts. Also, failures related to wheel tubes were excluded due to their high level of randomness.

MO data records specific dates for bike part repairs, however, this data structure is not optimal for conducting SA, which typically models the time to an event. To facilitate the use of SA, MOs were transformed into MO units. These units represent the time elapsed between two consecutive repairs for the same bike and bike component, and thus, it is the time period in which the bike part under study is operational. Consequently, when working with MO units, bike part survival information at the beginning and end of the data set could be lost for each bike. To prevent this information loss, the time from the start of the data set to the first MO is considered as one MO unit, and the time from the last repair of the bike to the end of the data set is regarded as another MO unit. It's worth noting that even though MO units may originate from the same bike, each one has been treated as an independent entity.

One key challenge in predicting the occurrence of an event is the presence of censored data, which refers to incomplete information about survival times \cite{leung1997censoring, gijbels2010censored}. Distinct types of censoring exist, including right-censoring (the event has not occurred by the end of the study), left-censoring (the event occurred before the study started), and interval-censoring (the event occurred within a specific time interval). Bicing maintenance data contains left- and right-censored data, since the first and last MO units of each bike and repair typology are incomplete. In the first unit, it is not possible to know when this bike part started working, and in the last, when this bike part finally will break. Since SA can effectively utilize uncensored and right-censored data to estimate the survival curves and generate predictions, only the left-censored MO units were discarded in the training and prediction phases. Specifically, this involves excluding the initial MO unit of each bike for the target bike part. Additionally, units without trips and the ones in which m-bikes were upgraded to e-bike were also discarded.

The data sets used for the survival models follow a format where each row encapsulates all the pertinent information about one subject. Within each entry, there are the details about the subject's survival duration, event occurrence, and aggregated covariate values crucial for the survival function estimation. MO units, with their defined start and end dates, facilitate the calculation of covariates that describe the utilization patterns of a bike part and its surrounding environmental factors. In this way, various covariates domains have been incorporated to the model inputs:

\begin{itemize}
\item \textbf{Weather}: weather data was analyzed to incorporate the surrounding environmental conditions of the MO unit. This involved considering the daily average temperature (in ºC), total daily precipitation (in mm), average wind direction (in degrees), mean wind speed (in km/h), and average atmospheric pressure (in hPa). After exploring this data and its connection to bike part failures, it was determined that the primary influential factors were the daily average temperature and the mean atmospheric pressure.
\item \textbf{Bike usage}: various metrics were computed, including daily distance traveled (in meters), daily positive and negative inclinations (in meters), and the mean daily speed (in km/h). These calculations were based on the previously determined shortest paths and inclinations between each couple of stations. Next, their cumulative versions were examined, and strong correlations were identified. As a result, only two variables were kept: the cumulative daily distance and the mean speed of the MO unit.
\item \textbf{Bike model}: due to potential variations in survival curves across the two bicycle models, a binary variable was introduced to distinguish between electric (1) and mechanical (0) bike models.
\item \textbf{Count of repairs for other bike parts during the target MO unit}: when analyzing a specific bike part, it becomes pertinent to evaluate the frequency of replacements for another bike part. This approach allows to gain valuable insights; for instance, understanding the number of wheel tubes replaced could aid in predicting potential replacements for components such as tires or wheel rims. Following an examination of the correlation and the variation inflation factor of these repair counts, the following MO subcategories were selected: brake tension adjustment, the replacement of the front and rear wheel tubes, and the replacement of the front wheel cover.
\end{itemize}

  \subsection{Models}
To predict the survival time of the bike components, the following statistical and machine learning models have been employed:

\begin{enumerate}
\item \textbf{Cox Proportional Hazard model (CPH)} \cite{cox1972regression} (lifelines' implementation \cite{lifelines}): is a semi-parametric approach that enables to evaluate how covariates influence the hazard rate of an event as time progresses. This model relies on an underlying assumption known as the proportional hazard assumption, which asserts that the relative risk between two distinct groups remains constant over time. Meeting this assumption simplifies the analytical process and enhances the interpretability of outcomes. Although evaluating the assumption holds theoretical importance for attaining a meaningful interpretation of covariates, adherence to this assumption might not be imperative. \cite{stensrud2020test} noted that when dealing with a sufficiently large sample size, even minor deviations from the assumption may show up. Furthermore, when the main goal is survival prediction, there is no need to test the proportional hazard assumption, since the main objective is to maximize an score \cite{lifelines_ph_assumption}. Consequently, for the purpose of this study, the analysis will prioritize predictive accuracy over strict adherence to the assumption.

\item \textbf{Multi-Task Logistic Regression model (MTLR)} \cite{yu2011learning} (pysurvival's implementation \cite{pysurvival}): stands as an alternative to CPH when the assumption of proportional hazards does not hold. MTLR model relies on a sequence of logistic regression models constructed across distinct time intervals. This allows the estimation of the probability associated with the occurrence of the event of interest within each interval. Consequently, the initial step involves specifying the desired number of intervals, with the present use case opting for the number of days in the maintenance data.

\item \textbf{Conditional Survival Forest model (CSF)} \cite{wright2017unbiased}  (pysurvival's implementation \cite{pysurvival}): is a machine learning model that extends Random Forest ensembles to effectively handle right-censored data. Therefore, this survival model allows to appropriately model data with non-linear relationships and censoring.

\item \textbf{CPH Deep Neural network model (DeepSurv)} \cite{katzman2018deepsurv}  (pysurvival's implementation \cite{pysurvival}): is an improved version of the CPH model that brings in elements of deep learning to its core structure. This enhancement enables the model to better capture complex patterns while still being able to handle censored data effectively.
\end{enumerate}

  \subsection{Hyper-parameters optimization}
The previous models possess distinct architectures. Consequently, the hyper-parameter optimization process has varied based on the specific model.

For the purpose of identifying the optimal hyper-parameters, the SA dataset of each bike component was partitioned into three subsets: training (60\%), validation (20\%), and test (20\%). Multiple hyper-parameter configurations were trained on the training set and assessed with the validation set and the root mean square error metric (RMSE) to identify the most favorable combination. Once determined, the model was trained using both the training and validation sets, and predictions based on the test set were generated to asses the final accuracy of the model. Furthermore, although model validation was primarily assessed using the RMSE metric, all data subsets were also evaluated for RMSE, determination coefficient ($R^2$) and mean absolute percentage error (MAPE) to assess accuracy comprehensively (Appendix \ref{app_accuracies}). RMSE and MAPE interpretation is very straightforward; the lower, the better. However $R^2$ works in the contrary direction since higher values, correspond to better predictions.

For the CPH model, the sole hyper-parameter subjected to optimization was the baseline estimation method (breslow, spline, or piecewise), while the remaining parameters were set to their default values. In the case of MTLR, CSF, and DeepSurv, the hyper-parameter search process was automated using the Optuna framework \cite{optuna_2019}. A total of 200 trials were conducted for each model, employing the TPEsampler class for hyper-parameter sampling and the MedianPruner class to halt unpromising combinations. For MTLR, the optimized hyper-parameters encompassed the learning rate (ranging between 1e-5 and 1e-3), initialization method (orthogonal or glotorot\_uniform), and optimizer (adam, adamaz, or sgd). The CSF optimization encompassed the number of trees (ranging between 10 and 100 in increments of 10), maximum tree depth (between 2 and 10), and minimum node size (ranging between 10 and 50 in steps of 5). In the case of DeepSurv, the optimization process considered initialization method (orthogonal or glotorot\_uniform), optimizer (sgd or adam), learning rate (ranging between 1e-5 and 1e-2), number of epochs (ranging between 50 and 500), L2 regularization (ranging between 0 and 1e-2), and the inclusion of batch normalization or dropout with a value of 0.5 (True or False for each one).

\section{Results}

\begin{figure}[!htbp]
  \centering
  \includegraphics[width=0.9\textwidth]{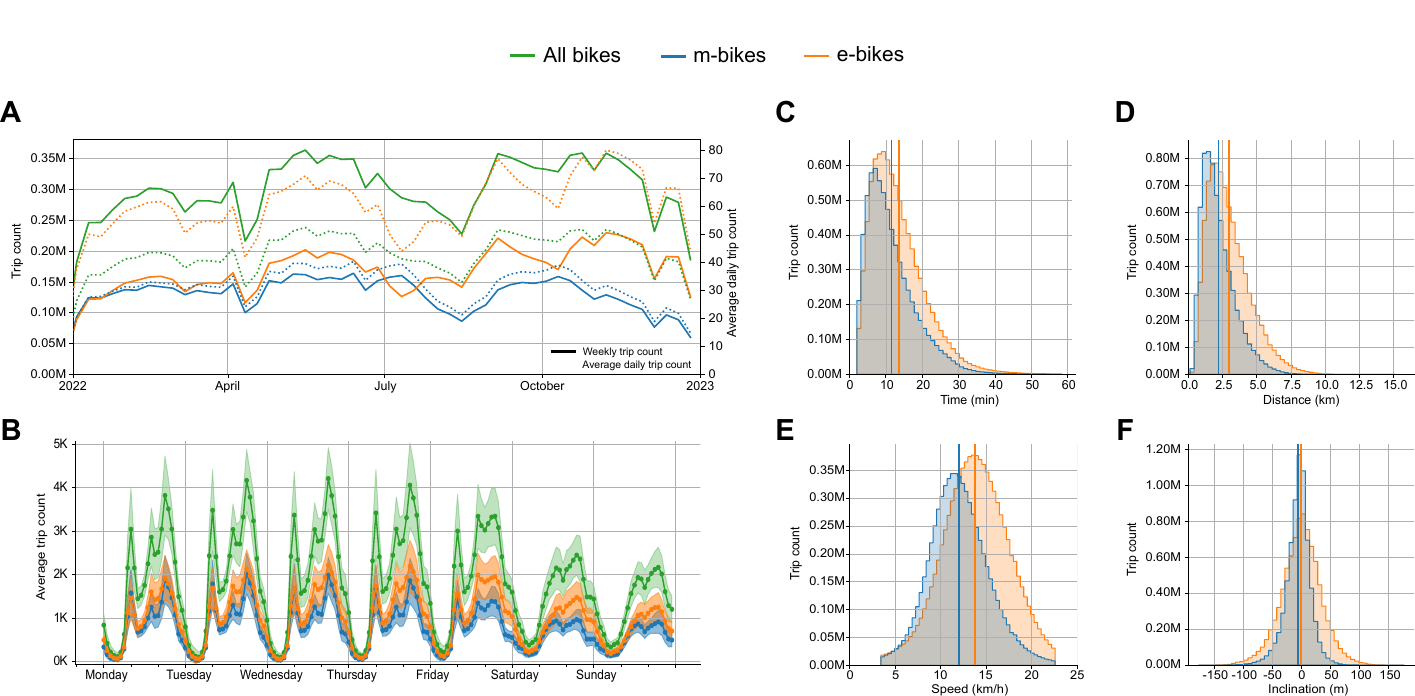}
  \caption{\textbf{Bike-sharing trip patterns in 2022.} (A) Time series depicting the weekly trip counts alongside the average daily number of trips per bike. (B) Hourly time series illustrating the average trip counts throughout the week, with the area around the time series indicating the standard deviation. (C-F) Distributions of trip characteristics, with vertical lines representing the mean of each subgroup.}\label{fig:trips_characteristics}
\end{figure}

The analysis of mobility patterns focused on data of 2022. Both mechanical and electric transportation modes exhibited variability across time due to seasonal variations and holidays such as Easter week and summer holidays (Figure \ref{fig:trips_characteristics}A). However, a noticeable differential trend emerged from summer on-wards. Electric mobility experienced a rise in trip numbers while the mechanical one suffered a significant decrease. Despite some m-bikes being upgraded to e-bikes in the course of 2022, the increase in the number of e-bike rides couldn't be entirely attributed to this transformation, considering they comprised only 40\% of the bike fleet at their highest point. E-bikes experienced a rise in the mean daily trips per bike, whereas m-bikes saw a significant reduction. Consequently, the variations in the usage of both transportation modes can be attributed to the upgrade of m-bikes into e-bikes and the increasing preference of users for the electric mobility.

  \subsection{Mobility patterns analysis} \label{section_results_mobility}

Trip counts present a strong weekly seasonality characterized by two distinct mainly daily patterns: weekdays and weekends (Figure \ref{fig:trips_characteristics}B). On weekdays, three important trip count maxima are observed at 8:00, 14:00, and 18:00, with a gradual increase in trip numbers as the week progresses, peaking on Thursday. Fridays slightly deviate from the typical weekday pattern since they present similar trip count peaks at 14:00 and 18:00. In contrast, weekends are characterized by a 30\% mobility reduction, the absence of an 8:00 peak, and a shift in the 18:00 peak to 19:00. Notably, electric mobility consistently matches or surpasses the mechanical one across all days and hours despite its additional cost. Both transportation modes also differ in their trip characteristics (Figure \ref{fig:trips_characteristics}C-F). Significant differences were found in the distributions of trip duration, distance, speed, and elevation (Section \ref{section_trips_processing}), implying two distinct behaviors. Specifically, electric mobility is characterized by steeper inclines, longer durations, greater distances from the origin, and higher speeds compared to mechanical mobility.

\begin{figure*}[!htbp]
  \centering
  \includegraphics[width=0.9\textwidth]{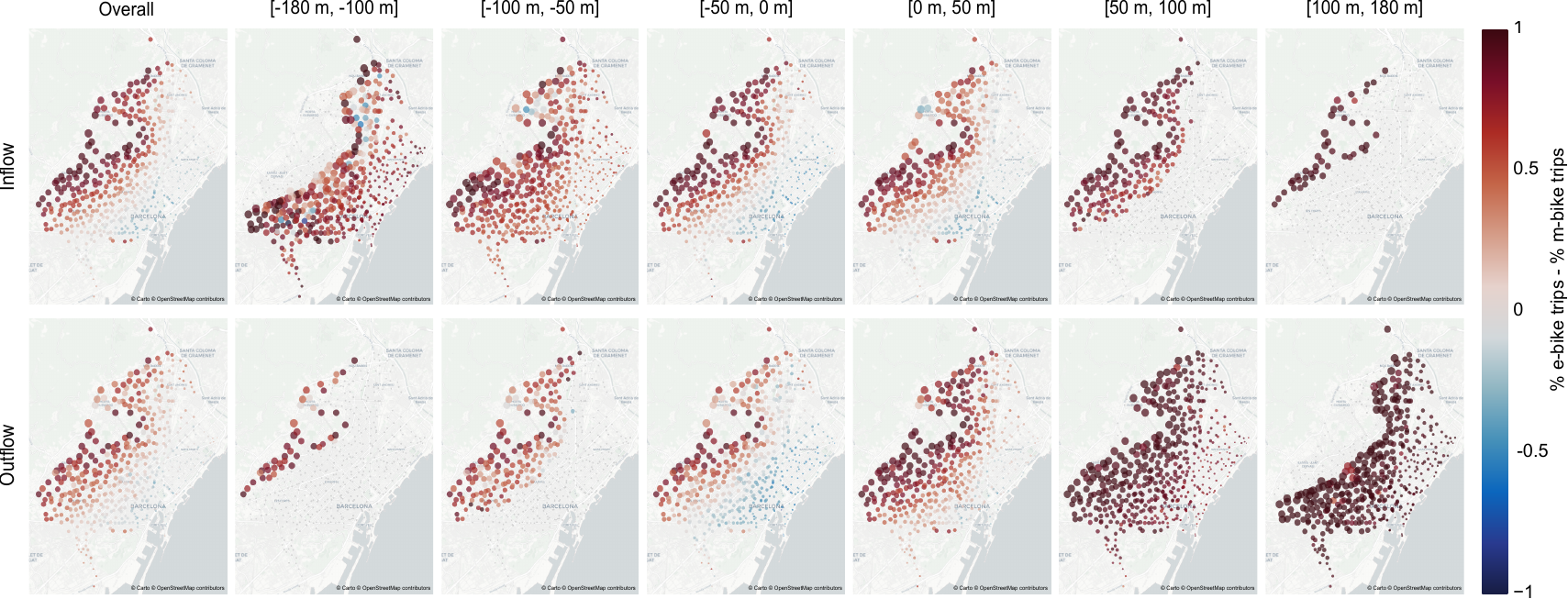}
  \caption{\textbf{Analysis of trip flow percentages for 2022, segmented by trip elevation.} Each pair of maps displays the differences in e-bike and m-bike trip percentages for incoming and outgoing trips at each station. The left maps include all trips, whereas the remaining figures apply a filter based on trip elevation. Within each map, individual dots represent BSS stations. The dot size reflects the station's altitude, with larger dots indicating higher altitudes, and the dot color signifies differences in trip flow percentages. Grey dots denote stations that do not receive or send trips with the specified altitude filter.}\label{fig:elevation_maps}
\end{figure*}

Additionally, each mobility mode exhibits unique spatial mobility dynamics (Figure \ref{fig:elevation_maps}). High-altitude stations predominantly receive e-bike trips, whereas low-altitude stations see the majority of incoming trips via m-bikes. This reflects users' preference for electric mobility when doing physically demanding trips. While this pattern is clear for incoming trips, it becomes less pronounced for outgoing trips, since the proportional utilization of e-bikes is not as dominant, even at high-altitude stations. This is attributed to the fact that the outgoing trips from a station do not indicate their destinations, resulting in a mix of trips to lower and higher stations. Nevertheless, given the overall preference for electric mobility among users, the inclination towards e-bikes still remains evident. Furthermore, when analyzing the differences in the percentages of incoming and outgoing trips involving e-bikes (Figure \ref{fig:sup:map_electric}), it becomes evident that the prevalence of incoming e-bike trips is more pronounced at high-altitude stations. In contrast, at lower-altitude stations, the percentages of incoming trips made by e-bikes become less substantial compared to the outgoing trips.

By segregating incoming and outgoing trips based on their elevation, this phenomenon becomes even more evident (Figure \ref{fig:elevation_maps}). At stations that generate outgoing trips with an elevation gain of over 100 meters, electric mobility completely dominates the transportation mode. Consequently, the receiving stations at higher elevations predominantly receive e-bike trips. This prevalence of electric mobility reduces gradually as the elevation decreases until reaching those trips with elevations between -50 and 50 meters. In this range of elevations, mechanical mobility plays a much more prominent role, especially at low altitude stations, however, even at high-altitude stations, electric mobility remains relevant in both incoming and outgoing trips. Notably, even in scenarios with negative elevations exceeding -50 meters, electric mobility continue to be the preferred transport mode choice, albeit with reduced percentages. This phenomenon may be attributed to users' preference for an electric transportation mode on uphill rides, resulting in more e-bikes accumulating at higher altitudes, and to potential elevation irregularities in paths between high-altitude stations.

  \subsection{Maintenance operations analysis}

Before generating survival predictions for bike components, an analysis of the failure patterns within the maintenance dataset was conducted. Given the low frequencies of most MO types and the criteria outlined in Section \ref{section_mos_processing}, only a few subcategories were deemed suitable for developing prediction models. Ultimately, three repair types were identified as the final targets for the predictions: brake pads, wheel spokes, and chains. Selected bike parts data was converted into MO units (Table \ref{tab:MO_units_censoring_vs_model}, Table \ref{tab:MO_units_cleaning}), and their corresponding covariate values were integrated as described in Section \ref{section_mos_processing}.

\begin{table}[!htbp]
  \centering
  \caption{\textbf{MO units: counts and percentages for the three chosen bike parts.}}
  \label{tab:MO_units_censoring_vs_model}
  \begin{tabular}{lccc}
      \toprule
                      & {M-bike}        & {E-bike}         & {Total}         \\
      \midrule
      \textbf{Brake pad MO units}                                           \\
      Uncensored     & 8,777 (19.4\%)  & 29,655 (65.5\%)  & 38,432 (84.9\%) \\
      Right-censored & 3,994 (8.8\%)   & 2,841 (6.3\%)    & 6,835 (15.1\%)  \\
      Total          & 12,771 (28.2\%) & 32,496 (71.2\%)  & 45,267 (100\%)  \\ \\
  
      \textbf{Wheel spokes MO units}                                        \\
      Uncensored     & 31 (0.2\%)      & 14,177 (83.6\%)  & 14,208 (83.8\%) \\
      Right-censored & 353 (2.1\%)     & 2,386 (14.1\%)   & 2,739  (16.2\%) \\
      Total          & 384  (2.3\%)    & 16,563  (97.7\%) & 16,947 (100\%)  \\ \\
  
      \textbf{Chain MO units}                                               \\
      Uncensored     & 4,344 (36.3\%)  & 1,679 (14\%)     & 6,023 (50.4\%)  \\
      Right-censored & 3,875 (32.4\%)  & 2,056 (17.2\%)   & 5,931 (49.6\%)  \\
      Total          & 8,219 (68.8\%)  & 3,735 (31.2\%)   & 11,954 (100\%)  \\
      \bottomrule
  \end{tabular}
  \end{table}

\begin{figure}[!htbp]
  \centering
  \includegraphics[width=0.9\textwidth]{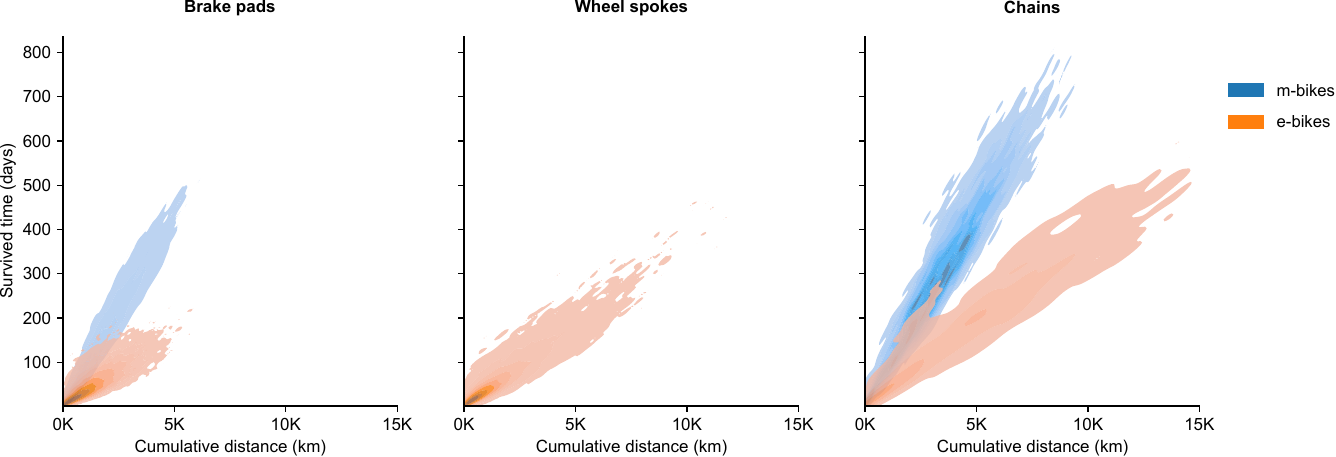}
  \caption{\textbf{Durability analysis of bike components.} This figure presents the distribution of the survival time for the uncensored MO units across various bike components, plotted against their cumulative traveled distance.}\label{fig:durability_analysis}
\end{figure}

The analysis of failure dynamics for the three bike parts revealed distinct patterns between m-bikes and e-bikes (Figure \ref{fig:durability_analysis}). Specifically, for brake pads and wheel spokes, e-bikes generally exhibit a significantly higher number of repairs per bike compared to their mechanical counterparts. Interestingly, m-bikes display very few wheel spoke repairs. Also, notable differences in the survival times of the three bike parts have been found. E-bike brake pads have shorter survival periods compared to their mechanical counterparts, and wheel spokes display similar trends. However, the scarcity of MOs for m-bikes makes direct comparisons of survival distributions impractical. Finally, chain repairs present the widest range of survival times for both m-bikes and e-bikes, with m-bike chains having longer survival times when compared to e-bike chains.

When exploring the potential relation with bike usage, cumulative distance emerged as the primary covariate that significantly distinguishes between bike models (Figure \ref{fig:durability_analysis}). M-bike brake pads demonstrate considerably longer durability compared to their e-bike counterparts when covering equivalent distances, while in the case of chains, the behavior is the opposite. Furthermore, although it appears that the limited number of uncensored m-bike MO units for wheel spokes outlast those of the e-bikes, the scarcity of the former prevents a meaningful comparison of their behaviors. Also, differences have been observed in terms of the average number of reparations for other MO subcategories during the target MO units (Table \ref{tab:other_mos_mean}). Thus, m-bikes and e-bikes bike parts exhibit distinct breakdown dynamics, which can be attributed to variations in how bikes are used based on their specific model.

\begin{figure}[!htbp]
  \centering
  \includegraphics[width=0.9\textwidth]{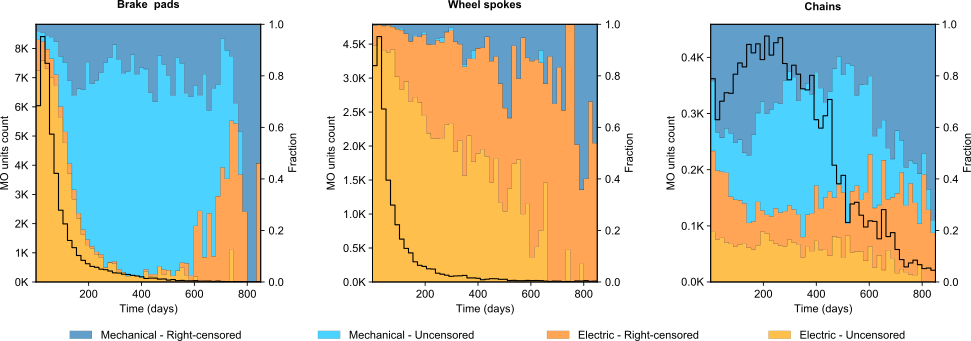}
  \caption{\textbf{Censoring distributions of bike components.} The graph features a black line representing the count of MO units over time. Accompanying this, colored regions illustrate the relative fractions of each MO unit type within the total count.}\label{fig:censoring}
\end{figure}

Before model training, MO units distributions according to censoring and bike model have been examined (Figure \ref{fig:censoring}, Table \ref{tab:MO_units_censoring_vs_model}). Uncensored data for both brake pads and wheel spokes show similar trends, representing in both cases approximately 84\% of the dataset. 90\% of the uncensored units have survival times of up to 200 days for brake pads and 150 days for wheel spokes, with some units lasting as long as approximately 800 days. However, there are two significant differences between both. First, the ratio of right-censored to uncensored units for brake pads remains fairly consistent until the 550-day mark. In contrast, for wheel spokes, the proportion of uncensored units steadily declines, virtually disappearing after 650 days. Secondly, uncensored units for m-bikes become the majority for brake pads after 200 days, whereas there is an almost complete absence of such units for m-bike wheel spokes. The decreasing number of MO units over time and the reduction in uncensored units suggest that modeling long-lasting MO units for both bike components could pose challenges. Moreover, the specific scarcity of uncensored MO units for m-bike wheel spokes in brake pads could introduce additional complexities in the modeling.

Chain maintenance exhibits distinct patterns. The number of censored units for chains exceeds that of the uncensored units, indicating that chains generally have longer lifespans compared to the other two bike components. Moreover, the failure rate for chains is notably different; only 30\% of chains fail before reaching the 200-day mark. Given the robust nature of chains, the proportion of uncensored units is low at the start of the timeline. Additionally, due to the rarity of instances surviving for extended periods, this proportion remains low past the 500-day mark as well. Considering the lower overall number of MO units for chains, coupled with the relatively smaller number of uncensored units and their reduced proportion at both the beginning and end of the timeline, modeling the longevity of chains could be more challenging compared to other parts.

  \subsection{Predictions accuracy}

Upon confirming data distributions in the training and test sets are aligned (Figure \ref{fig:sup:train_test_distributions}), survival models were trained and their performance was evaluated predicting the uncensored MO units from both datasets (Table \ref{tab:sup:prediction_results_train} and Table \ref{tab:prediction_results_test}). The decision to exclude the right-censored data was made to prevent comparisons of predictions with the time duration of MO units that do not conclude with a repair.

\begin{table}[!htbp]
  \centering
  \caption{\textbf{Predictions accuracy metrics on the test dataset.} Forecasts were generated using exclusively the uncensored MO units}
  \label{tab:prediction_results_test}
  \begin{tabular}{llccc}
      \toprule
                          & Models   & Brake pads     & Wheel spokes   & Chains         \\
      \midrule
      RMSE                 & CPH      & 66.27          & 91.07          & 100.64         \\
                          & MTLR     & 38.51          & 26.74          & 64.73          \\
                          & CSF      & 37.45          & 41.28          & 108.29         \\
                          & DeepSurv & \textbf{28.75} & \textbf{18.83} & \textbf{43.62} \\
      \midrule
      R\textsuperscript{2} & CPH      & 0.59           & -0.29          & 0.60           \\
                          & MTLR     & 0.86           & 0.89           & 0.84           \\
                          & CSF      & 0.87           & 0.73           & 0.55           \\
                          & DeepSurv & \textbf{0.92}  & \textbf{0.94}  & \textbf{0.93}  \\
      \midrule
      MAPE                 & CPH      & 59.65          & 68.96          & 197.59         \\
                          & MTLR     & 45.31          & 33.82          & 69.40          \\
                          & CSF      & 89.60          & 105.90         & 227.10         \\
                          & DeepSurv & \textbf{28.78} & \textbf{18.42} & \textbf{33.02} \\
      \bottomrule
  \end{tabular}
  \end{table}

As expected, across nearly all models, predictions made on a combination of the training and validation sets exhibit higher accuracy than those generated on the test set. This discrepancy arises because the training and the hyper-parameter optimization process has occurred on the training and validation sets, while the test set is strictly reserved for evaluating the final performance. Moreover, the close similarity between both sets' accuracies implies that the hyper-parameter optimization has been successful enough to grant the capacity for prediction generalization.

Classical CPH models have exhibited the poorest accuracy. This lesser performance is likely due to the their reliance on linear equations, which is inadequate for capturing the non-linear failure dynamics of the bike components. To address this issue, MTLR and CSF models were employed, demonstrating in both cases higher accuracies. However, an exception was noted with the CSF model's performance for chains. This outcome arises from the fact that CSF models were unable to generate predictions within the entire range of the training data, and thus, they were unable to adequately learn from the training set. In contrast, MTLR models clearly outperformed CPH models, indicating that the adapted logistic regressions for survival data are more effective than CPH models.

DeepSurv models were utilized to improve the modeling of non-linear data, achieving the best results. Compared to MTLR models, DeepSurv models demonstrated a reduction in RMSE values by 25 to 33\%. Also, they attained MAPE metrics below 33\% and R\textsuperscript{2} values exceeding 0.92. However, it is important to note that the accuracy of predictions for chains remains lower than those for brake pads and wheel spokes. This disparity might stem from two previously mentioned factors: the relatively small number of chain MO units and the limited availability of uncensored data for MO units with survival periods of either less than 200 days or more than 600 days.

  \subsection{Predictions analysis}

In the PM field, predictions must be as close as possible to the failure date and, ideally, these predictions should pertain to dates preceding the failure events, as it facilitates preventive maintenance and avoids the costs associated with late component replacements. In this line, an exploration of the predictions has been performed by considering right-censored and uncensored units separately.

Uncensored data predictions present a high level of accuracy. This assertion is further substantiated by a comparative analysis of the mean and standard deviations pertaining to the actual and predicted lifespans (Table \ref{tab:actual_vs_pred_descriptive}). However, right-censored MO units exhibit more substantial deviations in these statistics. Also, as previously stated, chain predictions exhibit the highest predictive errors, which can be attributed to its lower number of MO units and the significantly higher proportion of right-censored units.

\begin{table}[!htbp]
  \centering
  \caption{\textbf{Actual and predicted survival times descriptive table.} The means and standard deviations (in parenthesis) for the actual and predicted survival times (in days) are displayed.}
  \label{tab:actual_vs_pred_descriptive}
  \begin{tabular}{@{}lccc@{}}
      \toprule
                          & {All}           & {Uncensored}    & {Right-censored} \\
      \midrule
      \textbf{Brake pads MO units}                                              \\
      Actual survival    & 88.21 (96.95)   & 87.52 (97.83)   & 104.80 (103.81)  \\
      Predicted survival & 88.04 (91.19)   & 85.92 (91.86)   & 110.56 (92.42)   \\  \\
  
      \textbf{Wheel spokes MO units}                                            \\
      Actual survival    & 77.05 (88.58)   & 63.33 (71.73)   & 155.54 (126.64)  \\
      Predicted survival & 79.30 (91.62)   & 62.25 (69.58)   & 175.36 (130.84)  \\  \\
  
      \textbf{Chains MO units}                                                  \\
      Actual survival    & 258.70 (151.29) & 278.67 (147.79) & 230.81 (155.61)  \\
      Predicted survival & 283.02 (146.51) & 286.86 (145.69) & 272.73 (152.8)   \\
      \bottomrule
  \end{tabular}
\end{table}

\begin{figure}[!htbp]
  \centering
  \includegraphics[width=0.9\textwidth]{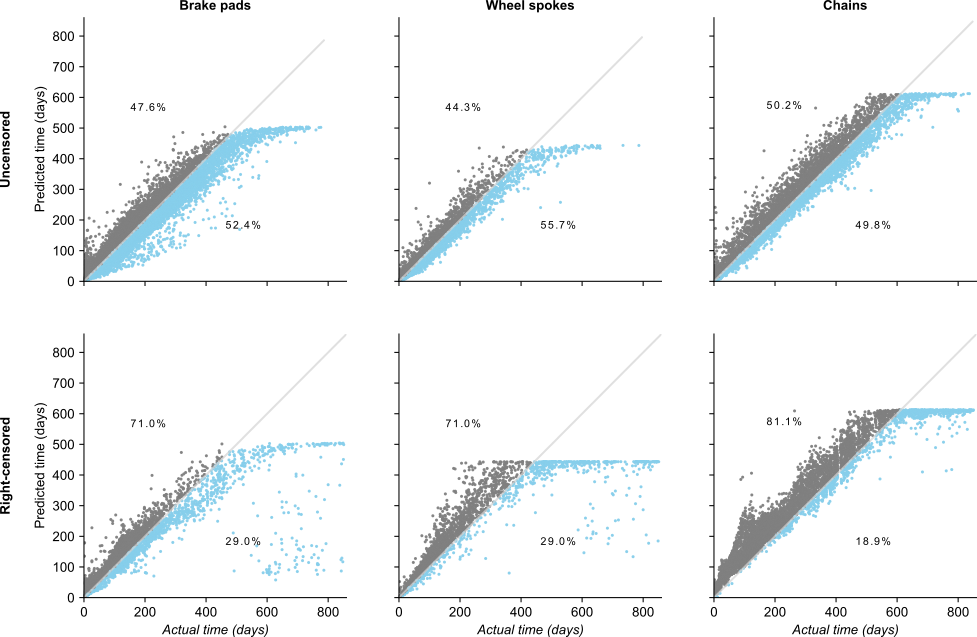}
  \caption{\textbf{Comparison of actual vs. predicted survival times of bike components.} Each plot displays instances predicted beyond the actual survival time in grey, and those predicted before in light blue. The charts also include percentages indicating the proportion of instances predicted to fail either before or after their actual failure dates}\label{fig:pred_exploration}
\end{figure}

Uncensored MO units exhibit strong lineal relationships between the predicted and actual survival times, as evidenced by Pearson correlation coefficients of 0.97 for brake pads, 0.97 for wheel spokes, and 0.96 for chains (Figure \ref{fig:pred_exploration}). Similarly, the predicted-to-actual survival time ratio for the three bike components is notably accurate (Table \ref{tab:pred_actual_statistics}), with a mean centered around 1, and approximately 80\% of predictions exhibiting an error of less than $\pm$25\% compared to the actual values. In terms of the absolute difference, in the brake pads and wheel spokes 80\% of the MO units have an error below 20 units, while for chains, it's 50\%. Therefore, it can be confidently concluded that the predictions are remarkably close to the actual failure dates.

Unlike the uncensored units, which provide data on when bike parts fail, the right-censored units lack information regarding the specific timing of these failures. For this reason, the predicted days are expected to have higher values than the actual days, and therefore the predicted/actual ratio values should tend values higher than one. In this case, indeed, the majority of predictions generated for the right-censored MO units were higher than the corresponding actual values (Figure \ref{fig:pred_exploration}).

\begin{table}[!htbp]
  \centering    
      \caption{\textbf{Descriptive statistics for predicted-actual value metrics of uncensored MO units.}}
      \label{tab:pred_actual_statistics}
      \begin{tabular}{@{}lcccccccccccccccc@{}}
      \toprule
      & mean  & std   & min  & 10\% & 20\% & 30\%  & 40\%  & 50\%  & 60\%  & 70\%  & 75\%  & 80\%  & 90\%  & 95\%  & 99\%   & max    \\
      \midrule
      \textbf{Predicted to actual ratio} \\
      Brake pads   & 1.09  & 0.76  & 0.25 & 0.76 & 0.84 & 0.89  & 0.94  & 0.99  & 1.04  & 1.11  & 1.14  & 1.19  & 1.34  & 1.57  & 3.38   & 40.51  \\
      Wheel spokes & 1.03  & 0.38  & 0.00 & 0.76 & 0.84 & 0.89  & 0.93  & 0.97  & 1.02  & 1.08  & 1.12  & 1.16  & 1.28  & 1.44  & 2.20   & 13.70  \\
      Chains       & 1.20  & 2.16  & 0.00 & 0.88 & 0.92 & 0.95  & 0.98  & 1.01  & 1.04  & 1.09  & 1.11  & 1.15  & 1.30  & 1.57  & 4.57   & 112.66 \\
      \\
      \textbf{Predicted minus actual} \\
      Brake pads   & 12.86 & 17.84 & 0.00 & 1.07 & 2.26 & 3.61  & 5.07  & 6.85  & 9.23  & 12.81 & 15.40 & 18.84 & 31.49 & 45.59 & 86.26  & 340.79 \\
      Wheel spokes & 8.75  & 12.42 & 0.00 & 0.74 & 1.54 & 2.44  & 3.52  & 4.83  & 6.52  & 8.71  & 10.30 & 12.37 & 20.37 & 30.52 & 61.60  & 223.53 \\
      Chains       & 28.05 & 25.81 & 0.01 & 3.76 & 7.66 & 11.78 & 16.28 & 21.13 & 26.69 & 34.40 & 38.63 & 44.33 & 61.66 & 77.47 & 115.11 & 334.97 \\
      \bottomrule
      \end{tabular}  
\end{table}

  \subsection{Models interpretability}

\begin{figure}[htbp!]
  \centering
  \includegraphics[width=0.9\textwidth]{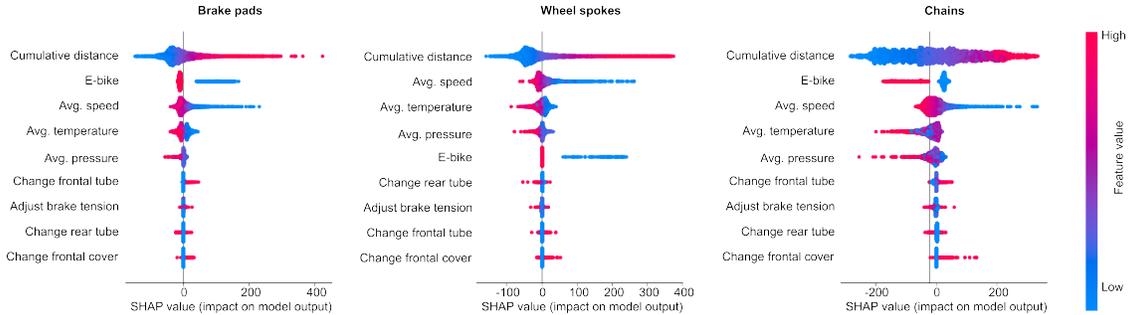}
  \caption{\textbf{SHAP values for survival time predictions.} SHAP values were calculated using 5000 samples from the training set and the KernelExplainer class. This explainer was created with 100 background representative samples from the training set, which were obtained using k-means. Features are ordered according to their mean absolute SHAP values in descending order.}\label{fig:shap}
\end{figure}

To assess how model inputs might affect the chances of models predicting a reparation, the game theoretic approach SHapley Additive exPlanations (SHAP) has been applied to the DeepSurv models of the three bike parts (Figure \ref{fig:shap}). Overall, the most influential features are the ones related to the bike usage and the bike model.

MO units cumulative distance emerged as the most influential factor in predicting the lifespans of the three bike parts. Larger values are strongly associated with longer predicted survival times. This connection is due to the fact that bike parts that survive for extended periods are able to cover significant distances. Therefore, the models consistently linked longer survival times with greater cumulative distances. Average mean speed have also emerged as one of the most critical features for all bike parts. Higher values are closely associated with shorter survival times, indicating that increased stress on the bike parts leads to a reduction in their lifespan.

In section \ref{section_results_mobility}, it was noted that the bike model plays an important role in BSS mobility dynamics. As a result, the bike model emerges as the second most impactful feature. Specifically, electric mobility is consistently linked with shorter survival times. However, it is worth noting that the impact of the bike model on wheel spokes is relatively less pronounced, which can be attributed to the fewer number of MO units for m-bikes.

While weather-related variables have a lesser impact, they remain significant. Higher mean daily temperatures and mean daily atmospheric pressure values were associated with reduced survival times for brake pads and wheel spokes. In essence, hotter and drier weather conditions tended to decrease the survival time of these bike components. However, this effect is less evident in the case of chains. This discrepancy can be attributed to their extended survival times, leading to a wider range of weather conditions being encompassed within the average values. Finally, the counts of repairs for other MOs have been found to be the variables with least impact in the predictions.

\section{Discussion}

This study offers a novel perspective in BSS research by analyzing mobility patterns, differentiating between m-bikes and e-bikes as separate modes of transportation. Our findings reveal that the Bicing BSS exhibits mobility dynamics similar to those in other BSS studies in terms of distance travelled, trip duration, and bike speed \cite{Jensen2010Characterizing, Ciancia2015Exploring, Zaltz2013Structure, Chiariotti2018Rebalancing, Borgnat2011SharedBI, Zhang2016Bicycle}. Notably, electric mobility has been found to be the preferred mode of transportation, characterized by longer trip durations, farther distances covered, and faster speeds when compared to mechanical mobility. Moreover, the trend of lower trip counts on weekends compared to weekdays, as noted in \cite{froehlich2009sensing, Bustamante2022Riding}, still holds true, along with the occurrence of three daily peaks \cite{froehlich2009sensing}. 

Aligned with the findings of \cite{Morency2015ModellingBU, Kim2020Anatomy}, our study further confirms that topography plays a significant role in BSS utilization. Electric mobility is primarily chosen for trips involving steep positive inclines, whereas mechanical mobility is preferred for routes with slight elevation changes, especially for journeys originating or ending at stations situated at lower elevations. Consequently, stations at higher altitudes predominantly receive e-bikes, whereas those at lower altitudes are more likely to attract the mechanical ones. Moreover, this pattern is even more pronounced when considering the elevation differences between the origin and ending stations. Interestingly, even in scenarios with significant negative elevations, where physical exertion is low, electric mobility remains the dominant choice. This preference may be attributed to the accumulation of e-bikes at higher altitudes and the irregularities in the paths connecting high-altitude stations.

Factors related to mobility, such as trip distance, not only differentiate electric from mechanical mobility but were also found to significantly influence the wear and tear on bike components. Leveraging these insights, in this paper we present a novel PM system for a BSS, which marks a significant step towards replacing corrective maintenance strategies. The system has been designed to predict maintenance needs for three essential bike components, delivering satisfactory forecasting results through the application of deep learning survival models. The design we introduce enables operation across an entire bike fleet, distinguishing it from previous research in bike PM that primarily focused on individual bikes \cite{MountainBike2019, Matkovic2021Towards}. Moreover, our system enhances upon key aspects of an earlier PM solution developed for Oslo's BSS \cite{Predictivemaintenancebike}. First, our datasets are considerably more comprehensive. Second, we treat different repair typologies independently, acknowledging their distinct breakdown dynamics.  Lastly, to prevent potentially discriminatory conclusions that could impact BSS pricing strategies, we have deliberately omitted user information such as gender and age from our modelling. Additionally, through the application of a game-theoretic interpretability approach, we verify that the model's predictions are consistent with the observed failure dynamics. Notably, the most influential factors in the generation of the predictions are the cumulative distance and whether the bike is mechanical or electric.

The results of this study, along with their potential implications, hold significant promise for impacting society, especially in terms of enhancing the sustainability of urban mobility. Understanding BSS users' preferences for mechanical and electric mobility is essential for improving the decision-support systems that facilitate fleet rebalancing. Enhancing these systems not only assists BSS managers but also promotes BSS mobility by improving the user’s experience, ensuring the availability of the preferred mode of transport when needed. On the other hand, the implementation of PM systems also promotes a more sustainable mobility. Their potential to considerably reduce the environmental footprint and operational costs of BSS through more efficient resource utilization is remarkable. Furthermore, these systems enhance the user’s experience by ensuring a more reliable and available bike fleet. Despite facing challenges such as unbalanced datasets and the necessity of relying on approximations for bike usage inference, this study demonstrates the viability of deploying a successful PM system. Nevertheless, employing interpretability tools to establish trust in the PM systems' accuracy among BSS managers is essential for their adoption.

\section*{List of abbreviations}

\begin{itemize}
    \item \textbf{BSS}: bike-sharing system
    \item \textbf{e-bike}: electric bike
    \item \textbf{m-bike}: mechanical bike
    \item \textbf{MO}: maintenance operations
    \item \textbf{ARMA}: auto-regressive moving average
    \item \textbf{PM}: predictive maintenance
    \item \textbf{SA}: survival analysis
    \item \textbf{CPH}: Cox proportional hazard model
    \item \textbf{MTLR}: multi-task logistic regression model
    \item \textbf{CSF}: conditional survival forest model
    \item \textbf{DeepSurv}: CPH deep neural network model
    \item \textbf{SHAP}: shapley additive explanations
    \item \textbf{RMSE}: root mean square error
    \item \textbf{MAPE}: mean absolute percentage Error
    \item \textbf{$R^{2}$}: determination coefficient
\end{itemize}

\clearpage

\section*{Availability of data and materials}

The data that support the findings of this study are available from Pedalem-Bicing but restrictions apply to the availability of these data, which were used under license for the current study, and so are not publicly available.

\section*{Competing interests}

The authors declare that they have no competing interests.

\section*{Funding}

This work was supported through funds granted by the Government of Catalonia in the frame of the CATALONIA.AI programme. JGE is a fellow of Eurecat's "Vicente López" PhD grant program.

\section*{Authors' contribution}

JGE, AB and JV conceptualised the study. JGE developed the methodology, performed analyses and models. JGE wrote the original draft; and all authors critically discussed the results, revised the paper and approved the final manuscript.

\section*{Acknowledgements}

We thank Daniel Santanach, coordinator of CATALONIA.AI programme. We thank Marco Orellana from CIDAI (Centre of Innovation for Data Tech and Artificial Intelligence). We want to thank Faustino Corchero from Barcelona de Serveis Municipals and Roger Junqueras, Irene Giménez and all collaborators from Pedalem-Bicing. We also thank Javier Bejar from UPC-IDEAI.

\bibliography{sn-bibliography}

\clearpage

\begin{appendices}

\section{Data exploration} \label{app_data}
\setcounter{figure}{0}
\setcounter{table}{0}
\renewcommand{\thefigure}{A\arabic{figure}} 
\renewcommand{\thetable}{A\arabic{table}} 

    \begin{figure}[htpb!]
        \includegraphics[width=0.9\textwidth]{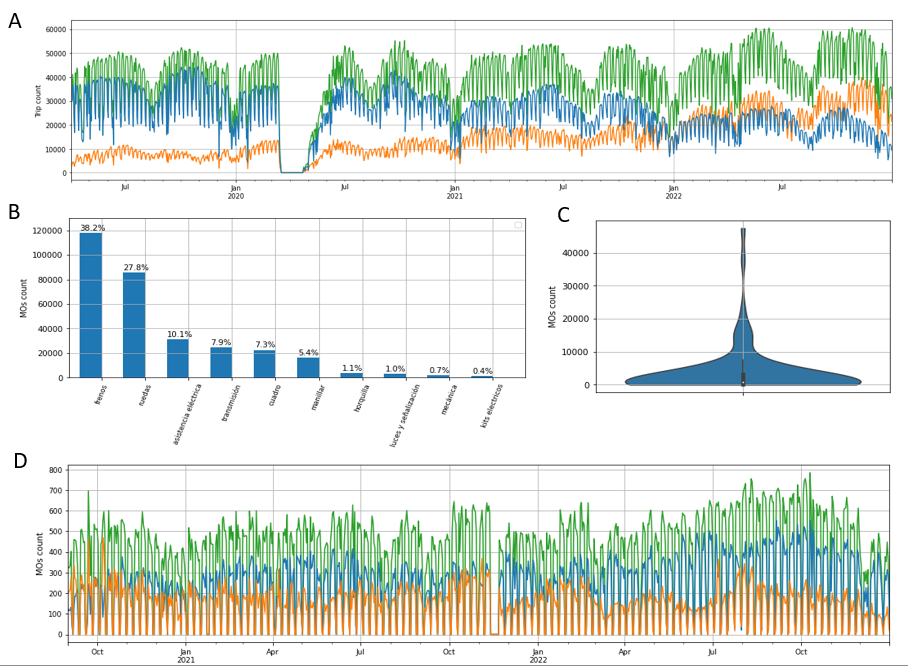}
        \caption{\textbf{Data overview.} (A) Time series of daily bike ride counts. The green line depicts all trips, the blue line represent the subset of trips completed on m-bikes, and the orange lines represent the subset of trips conducted on e-bikes. (B) Quantitative analysis of MOs with counts and percentages across all categories. (C) Distribution of MOs within subcategories. (D) Time series of daily MO counts.}
        \label{fig:supp:data_section}
    \end{figure}
    
\clearpage
\section{Mobility analysis}
\setcounter{figure}{0}
\setcounter{table}{0}
\renewcommand{\thefigure}{B\arabic{figure}} 
\renewcommand{\thetable}{B\arabic{table}} 

    \begin{figure}[htpb!]
        \centering
        \includegraphics[width=0.7\textwidth]{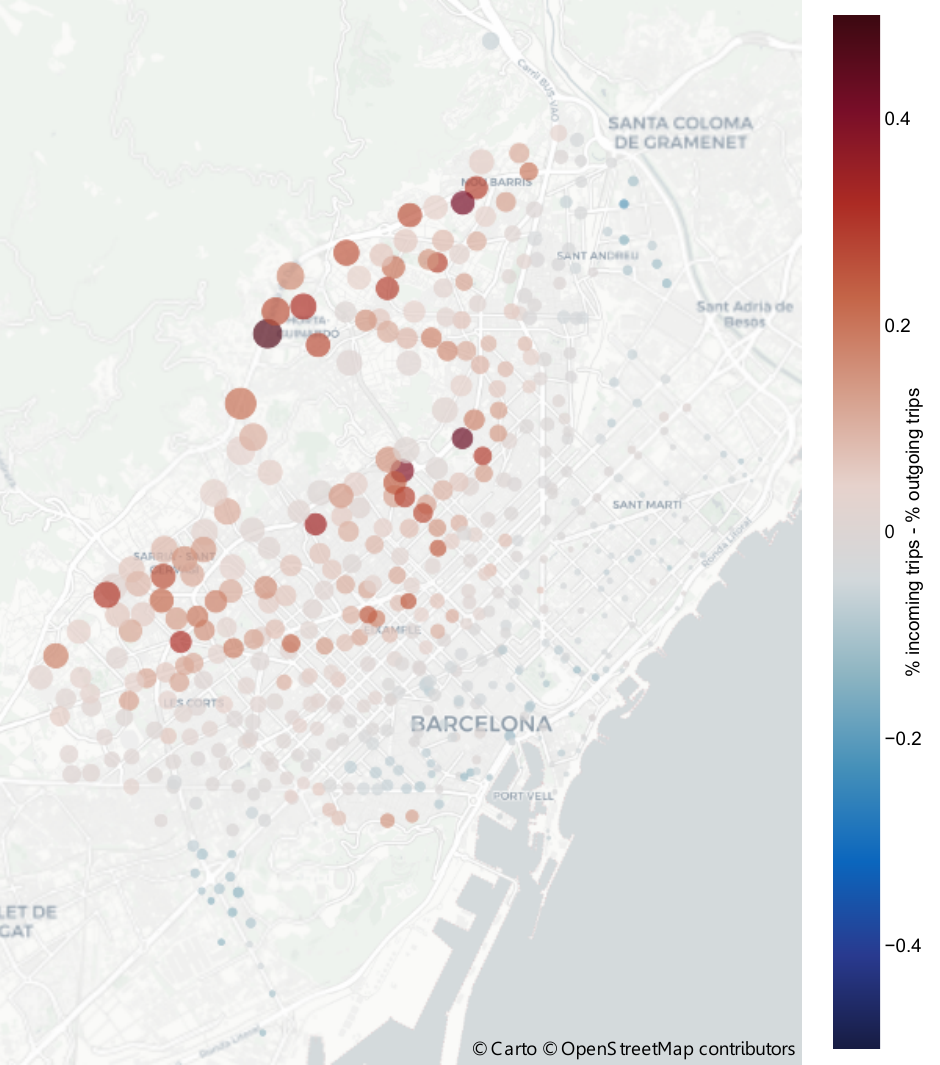}
        \caption{\textbf{Analysis of E-bike Trip Flow Percentages in 2022.} The map highlights the differences in incoming and outgoing trip percentages for e-bikes, noting that these differences are the inverse of those observed for m-bikes. Each dot represents a station of the BSS. The dot size reflects the station’s altitude, with larger dots indicating higher altitudes, and the dot color signifies differences in trip flow percentages.}
        \label{fig:sup:map_electric}
    \end{figure}

    \clearpage

\section{Maintenance operations analysis}
\setcounter{figure}{0}
\setcounter{table}{0} 
\renewcommand{\thefigure}{C\arabic{figure}} 
\renewcommand{\thetable}{C\arabic{table}} 

    \begin{table}[htpb!]
        \centering
        \caption{\textbf{Count of MOs and MO units.} The first three columns contains the MO count for all bikes and categorized by bike model. The fourth column, the total count of MO units. Since each bike has one more MO unit than its number of repairs, the amount of MOs for a bike part is less than its total number of MO units. The fifth column displays the count of MO units after excluding left-censored units, units without associated trips and units where the bike model changes. The sixth and seventh columns categorize the remaining MO units into m-bike and e-bike MO units, and the last two columns, into right-censored and uncensored MO units.}
        \label{tab:MO_units_cleaning}
        \begin{tabular}{lccc}
            \hline
            & Brake pads & Wheel spokes & Chains \\
            \hline
            MOs & 47,513 & 17,377 & 13,019 \\
            M-bike MOs & 14,809 & 516 & 9,433 \\
            E-bike MOs & 32,704 & 16,961 & 3,586 \\
            \hline
            MO units & 53,103 & 23,941 & 19,682 \\
            Clean MO units & 45,267 & 16,947 & 11,954 \\
            \hline
            M-bike MO units & 12,771 & 384 & 8,219 \\
            E-bike MO units & 32,496 & 16,563 & 3,735 \\
            \hline
            Uncensored MO units & 38,432 & 14,208 & 6,023 \\
            Right-censored MO units & 6,835 & 2,739 & 5,931 \\
            \hline
    \end{tabular}
    \end{table}

    \begin{table}[htpb!]
        \centering
            \caption{\textbf{Mean number of reparations and standard deviations (in parentheses) for the repair typologies within the target MO units.} A statistical comparison of means is conducted using the Welch t-test in cases of unequal variances and the t-test in cases of equal variances. Only the repair typologies included as model inputs are presented in the table.}
        \label{tab:other_mos_mean}
        \begin{tabular}{@{}lccc@{}}
        \hline
            &  M-bike & E-bike & P-value\\
        \hline
        \textbf{Brake pads MO units}\\
            Brake tension adjustment & 0.84 (1.08) & 0.65 (0.99) & 0.000 \\
            Front cover change &  0.03 (0.18) &  0.03 (0.16) & 0.000 \\
            Frontal tube change &  0.71 (0.97) &  0.26 (0.59) & 0.000\\
            Rear tube change &  0.13 (0.46) &  0.23 (0.68) & 0.000\\
        \hline
        \textbf{Wheel spokes MO units}\\
            Brake tension adjustment & 1.00 (1.39) & 0.96 (1.42) & 0.000 \\
            Front cover change & 0.00 (0.00) & 0.03 (0.18) & 0.000 \\
            Front tube change & 0.68 (0.94) & 0.34 (0.82) & 0.000 \\
            Rear tube change & 0.10 (0.40) & 0.35 (0.94) & 0.000 \\
        \hline
        \textbf{Chains MO units}\\
            Brake tension adjustment & 1.26 (1.28) & 3.48 (3.00) & 0.000\\
            Front cover change & 0.04 (0.19) & 0.10 (0.32) & 0.000 \\
            Front tube change & 1.01 (1.19) & 1.28 (1.64) & 0.000 \\
            Rear tube change & 0.21 ( 0.57) & 1.19 (1.94) & 0.000 \\
        \hline
        \end{tabular}
    \end{table}

    \clearpage
\section{Accuracy measures} \label{app_accuracies}
    To compare the accuracy of the predictive models, three metrics have been used:
    \begin{itemize}
    \item Root Mean Square Error (\textbf{RMSE}): is a non-proportional metric with the same units as the target variable that accentuates the effect of the time-steps with big forecasting error.
    \item Mean Absolute Percentage Error (\textbf{MAPE}): is a proportional metric in percentage units that represents the average absolute deviation between the predicted and the actual values.
    \item Determination coefficient (\textbf{R\textsuperscript{2}}): represents the proportion of the variance in a dependent variable (actual values) that is predictable from the independent variables (predicted values). It ranges from 0 to 1, where 0 indicates that the model does not explain any variance in the data, and 1 indicates that the model perfectly fits the data.
    \end{itemize}
    
    All of them are described in the following equations, where $y$ is the actual days that a bike part survives, $\hat{y}$ is the forecasted, and $n$ is the amount of MO units for a particular bike part:

    $$ RMSE = \sqrt{\frac{1}{n}\sum_{i=1}^{n}(y_i - \hat{y}_i)^2} $$

    $$ R^2 = 1 - \frac{\sum_{i=1}^{n}(y_i - \hat{y}_i)^2}{\sum_{i=1}^{n}(y_i - \bar{y})^2} $$

    $$ MAPE = \frac{1}{n} \sum_{i=1}^{n} \left| \frac{y_i - \hat{y}_i}{y_i} \right| \times 100 $$
    
    To sum up, RMSE and MAPE interpretation is very straightforward; the lower, the better.  However $R^{2}$ works in the contrary direction since higher values, correspond to better predictions.

    \clearpage

\section{Model training}
\setcounter{figure}{0}
\setcounter{table}{0} 
\renewcommand{\thefigure}{E\arabic{figure}} 
\renewcommand{\thetable}{E\arabic{table}} 

        \begin{figure*}[htpb!]
            \centering
            \includegraphics[width=0.9\textwidth]{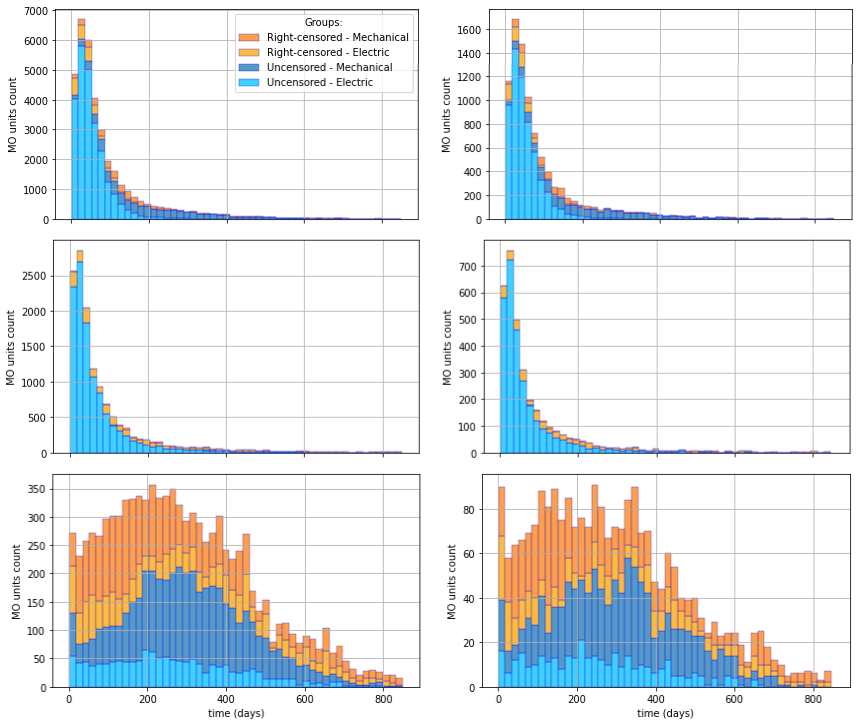}
            \caption{\textbf{Training and test sets distributions for the survival time.} Survival distributions for the brake pads (top), wheel spokes (middle) and chains (bottom) MOs. The training set encompasses both the training and validation data sets.}
            \label{fig:sup:train_test_distributions}
        \end{figure*}
        
        \begin{table*}[htpb!]
            \centering
            \caption{\textbf{Predictions accuracy metrics on the training dataset.} Forecasts were generated using exclusively uncensored MO units. The training set includes both the training and validation datasets.}
            \label{tab:sup:prediction_results_train}
            \begin{tabular}{llccc}
            \hline
            & Models & Brake pads & Wheel spokes & Chains \\
            \hline
            RMSE & CPH & 62.60 & 93.32 & 101.60 \\
            & MTLR & 33.50 & 27.32 & 63.21 \\
            & CSF & 35.48 & 37.08 & 103.77 \\
            & DeepSurv & \textbf{27.46} & \textbf{17.87} & \textbf{40.62} \\
            \hline
            R\textsuperscript{2} & CPH & 0.61 & -0.32 & 0.59 \\
            & MTLR & 0.89 & 0.89 & 0.84 \\
            & CSF & 0.87 & 0.79 & 0.57 \\
            & DeepSurv & \textbf{0.92} & \textbf{0.95} & \textbf{0.93} \\
            \hline
            MAPE & CPH  & 55.40 & 73.11 & 158.80 \\
            & MTLR & 36.54 & 38.53 & 60.28 \\
            & CSF & 83.15 & 106.50 & 147.76 \\
            & DeepSurv & \textbf{26.87} & \textbf{18.76} & \textbf{26.48} \\
            \hline
            \end{tabular}
        \end{table*}

\end{appendices}

\end{document}